# Magnetic and Electronic properties of $Eu_{0.9}Ca_{0.1}BaCo_2O_{5.5+\delta}$ with the disparity of Oxygen Stoichiometry

Md. Motin Seikh[1,2,*], Asish K. Kundu[1,3], V. Caignaert[1], V. Pralong[1] and B. Raveau[1]

[1]*CRISMAT, ENSICAEN-CNRS UMR 6508, 6 bd. Maréchal Juin, 14050 Caen, FRANCE*
[2]*Department of Chemistry, Visva-Bharati University, Santiniketan-731235, India*
[3]*Indian Institute of Information Technology Design & Manufacturing, Dumna Airport Road, Jabalpur –482005, India*

## Abstract

The effect of oxygen content on the magnetic and transport properties of the ferromagnetic $Eu_{0.9}Ca_{0.1}BaCo_2O_{5.5+\delta}$ has been carried out. Unlike the increase in $T_C$ with calcium content, paradoxally the $T_C$ value decreases with the increase in oxygen ($Co^{4+}$) content as observed in the undoped phase. This result unveils the hidden generic magnetic feature of the $LnBaCo_2O_{5.5}$ system in the calcium doped phase. This behaviour supports strongly the appearance of $Co^{3+}$ disproportion action into $Co^{4+}$ and $Co^{2+}$ and the magnetic phase separation scenario of ferromagnetic domains embedded in an antiferromagnetic matrix. All the samples covering a wide range of oxygen content, exhibit a p-type conductivity.

[*] Corresponding author: mdmotinseikh@yahoo.com (M.M.Seikh), Tel:+91-03463-261526

# Introduction

Transition metal oxides with the perovskite structure exhibit a wide range of fascinating physical properties due to the strong interplay between spin, charge, orbital and lattice degrees of freedom. However, the cobaltites display an additional parameter, namely spin state transition which throws an extra challenge to understand magnetism in cobalt oxides. Like other transition metal oxides, cobaltites can accommodate oxygen vacancies, giving rise to variable ratios of $Co^{2+}$, $Co^{3+}$ and $Co^{4+}$, each with three possible spin states: the low-spin (LS), the intermediate-spin (IS) and the high-spin (HS) states, which have a strong influence on the electrical and magnetic properties on the system.

The ordered oxygen deficient cobaltites with the formula $LnBaCo_2O_{5.50}$, have been the matter of numerous investigations, due to their fascinating magnetic and transport properties [1-23]. The crystal structure of these oxides can be described as the stacking sequence of $[CoO_2]$-$[BaO]$-$[CoO_2]$-$[LnO_\delta]$ along the **c** direction, where the oxygen vacancies are adopted in the $[LnO_\delta]$ planes [3,16]. Besides the existence of magnetoresistance and metal-insulator transition ($T_{IM}$) at room temperature, these cobaltites commonly exhibit successive paramagnetic-ferromagnetic-antiferromagnetic (PM-FM-AF) transitions [1-3]. There are various models which have been proposed to explain such complex magnetic transitions [6,16,18-20]. The spin-state of cobalt ions is also a standing issue in this layered cobaltite.

Recently, we have reported that the substitution of lanthanide by calcium leads to a dramatic change, especially in the magnetic properties [24-26]. When calcium is substituted for the smaller size lanthanides the ferromagnetic state is destroyed [26]. However, for the larger size lanthanide, it leads to the expansion of the ferromagnetic state at the expense of the antiferromagnetic phase. Interestingly, this change in magnetic properties occurs for a cobalt valence close to ~+3. It was also shown that the origin of the ferromagnetic phase in the pure and doped compositions is distinct in nature [24-25]. The present work is focused on the effect of oxygen content in a calcium doped ferromagnetic phase $Eu_{0.9}Ca_{0.1}BaCo_2O_{5.5+\delta}$. We show that, though Ca doping stabilizes the ferromagnetic state through $Co^{3+}$ disproportionation (formation of $Co^{4+}$ and $Co^{2+)}$, paradoxally $T_C$ decreases with the increase of oxygen content, i.e. with increasing $Co^{4+}$ content. This behaviour, similar to that previously observed for the undoped $LnBaCo_2O_{5.5+\delta}$ phases, supports strongly the phase separation scenario of FM domains embedded in the AFM matrix, suggesting that the oxidation will mainly affect the AFM matrix.

**Experimental**

The samples $Eu_{0.9}Ca_{0.1}BaCo_2O_{5.5+\delta}$, were synthesized from mixtures of oxides $Eu_2O_3$, $CaCO_3$, $BaCO_3$ and $Co_3O_4$ in stoichiometric ratio by standard ceramics method. $Eu_2O_3$ was dried at 900 °C prior mixing with $BaCO_3$, $CaCO_3$ and $Co_3O_4$ used as received. The appropriate proportions of the staring materials were weighed and thoroughly mixed by mortar pestle adding ethanol for homogeneous mixing. The intimate mixture was first heated in air at 1000 °C for 24 h. They were then ground and pressed in the form of bars and sintered in air at 1100 °C for 24 h and finally cooled down to room temperature at 100 °C/h. The X-ray diffraction (XRD) patterns were recorded using a Philips X'pertPro diffractometer with Cu Kα radiation. The energy-dispersive spectroscopy (EDS) analysis performed with a kevex analyzer mounted on a JEOL 200 CX electron microscope allowed the normal cationic composition to be confirmed. The oxygen content determined by iodometric titration in argon atmosphere allowed the sample formula to be determined as $Eu_{0.9}Ca_{0.1}BaCo_2O_{5.48}$ ($\delta$ = -0.02). Though the change in oxygen stoichiometry is reversible in these cobaltites, to achieve various oxygen contents samples we have treated the separate bars at different conditions in argon and oxygen atmospheres. The samples annealed in argon atmosphere at 550 and 400 °C for 12 h exhibit the compositions $Eu_{0.9}Ca_{0.1}BaCo_2O_{5.22}$ ($\delta$ = -0.28) and $Eu_{0.9}Ca_{0.1}BaCo_2O_{5.43}$ ($\delta$ = -0.07), respectively. Whereas the oxygen annealing of the samples at 50 and 100 bars of pressure at 600 °C for 12 h gives the compositions $Eu_{0.9}Ca_{0.1}BaCo_2O_{5.60}$ ($\delta$ = 0.10) and $Eu_{0.9}Ca_{0.1}BaCo_2O_{5.63}$ ($\delta$ = 0.13), respectively, determined by iodometric titration within an error of ±0.02.

Electrical indium contacts were ultrasonically deposited on bars with typical dimensions 2 × 2 × 10 mm$^3$ to perform electrical measurements. The resistance was measured by standard four-probe technique within the temperature range of 5 to 400 K and thermoelectric power (steady-state method) within the temperature range of 5 to 320 K. The measurements were carried out in a Quantum Design Physical Properties measurements system (PPMS). The magnetic measurements were carried out with a SQUID magnetometer (MPMS, Quantum Design). For each sample, the zero field-cooled (ZFC) and field-cooled (FC) data were collected with an applied field of 100 Oe. The magnetization versus applied field (M-H) measurements were performed with an applied field of ±5 Tesla.

**Results**

Fig. 1 shows the room temperature X-ray diffraction (XRD) patterns for all the compositions of $Eu_{0.9}Ca_{0.1}BaCo_2O_{5.5+\delta}$. All the XRD patterns can be indexed in an orthorhombic P*mmm* space group. In Fig. 2, we have shown the variation of the lattice parameters, as a function of oxygen content, $\delta$. The doubling of the cell parameters takes place along the b and c directions giving rise to the so-called $a_p$ x $2a_p$ x $2a_p$ unit cell, characteristic of the "112" type structure ($a_p$ is the perovskite cell parameter [3]). There are significant changes in the lattice parameters with the variation in oxygen content. The **a** parameter shows little increase with the increase of oxygen content till $\delta = -0.02$ followed by sudden a decrease for higher $\delta$ values as shown in Fig. 2. The **b** and **c** parameters show little contraction and expansion, respectively, with the increase in oxygen content (Fig. 2). However, as expected, the cell volume decreases with the increase in oxygen content (see inset of Fig. 2), in agreement with the increasing content of $Co^{4+}$, of smaller size than $Co^{3+}$.

In Fig. 3, we present the zero filed cooled (ZFC) and field cooled (FC) magnetization curves of $Eu_{0.9}Ca_{0.1}BaCo_2O_{5.5+\delta}$ registered in an applied field of 100 Oe. The $\delta = -0.02$ composition exhibits a jump in magnetization at ~310 K signifying the appearance FM state (Fig. 3(c)). This ferromagnetic ordering temperature is significantly higher than that of the stoichiometric $EuBaCo_2O_{5.5}$ phase [3,7]. However, one can observe a small kink immediately below the $T_C$ and a small hump on further lowering the temperature. The sample with $\delta = -0.28$ does not show any prominent FM ordering, except certain change in slope of the magnetization curve, in the temperature range 5-400

K and the magnetization value is very low (Fig. 3(a)) compared to others. However, we noticed two clear discontinuities around 320 and 100 K (Fig. 3(a)). The slope change in the M(T) curve at ~320 K (Fig. 3(a)) is related to the generic metal-insulator transition exhibited by this family of cobaltites [3]. The hump near 100 K could be associated with the PM-AF transition. The $\delta$ = -0.07 composition shows a clear ferromagnetic transition around 300 K and a huge thermomagnetic hysteresis below $T_C$ (Fig. 3(b)). For higher values of $\delta$ (0.10 and 0.13) the ZFC-FC magnetization curves show a clear PM-FM transition followed by a sharp peak related to FM-AF transition (Fig. 3(d)) and on further lowering temperature the FC magnetization increases (though ZFC magnetization remains constant). However, the $T_C$ values are significantly lower than for other compositions (Fig. 3(d)). The increase in magnetization (see Fig. 3) on lowering the temperatures could be related to the growth of ferromagnetic domains which are embedded inside the antiferromagnetic matrix. The inset in Fig. 3(b) shows the variation of $T_C$ with oxygen content. The $T_C$ value decreases with the increase in the increase in cobalt valence.

We have registered isothermal magnetization, M(H) curves at different temperatures to confirm the nature of magnetic ordering. Fig. 4 shows the M(H) curves recorded at 10 and 100 K for all the compositions. The M(H) data undergo dramatic changes in the magnetically ordered state with oxygen disparity. The samples for $\delta$ = -0.28 and –0.07 do not show any hysteresis loop at 10 K (Fig 4(a)), implying the absence of the FM ordering. All other compositions show clear hysteresis loop with large coercive field (Fig. 4(a)). Nevertheless, the field-induced magnetic transition is weak and significantly broadened, a continuous increase in slope in M(H) data beyond 16 kOe is observed in the increasing field direction. On the other hand, except $\delta = -0.28$ composition, all other samples show typical hysteresis loop at 100 K, which is a clear signature of FM ground state (Fig. 4(b)). For higher oxygen content, the variation of M with H in the magnetically ordered state reveals the similar kind of spin orientation effects at both temperatures. It is important to note that, for $\delta > -0.02$, the value of magnetic moment at 100 K is higher than at 10 K (for H > 4 kOe). To further characterize the low temperature complex magnetic ordering we have carried out frequency dependence magnetic ac-susceptibility measurements.

Fig. 5 shows the temperature dependence of ac-susceptibility for $Eu_{0.9}Ca_{0.1}BaCo_2O_{5.63}$ at four different frequencies. The in phase component $\chi'(T)$ of the ac-susceptibility reveals similar features as the ZFC-magnetization in low applied DC field (100 Oe). The sample exhibits a sharp maximum corresponding to their FM $T_C$, which is frequency dependent and shifting towards higher temperature with increasing frequencies. The out phase component $\chi''(T)$, also exhibits similar frequency dependent behavior as shown in the inset of Fig. 5. In addition, we noticed prominent frequency dependence just below the peak position. A similar behaviour was reported earlier in 112-ordered $LaBaCo_2O_6$ cobaltite and explained as glassy ferromagnetism [27]. For the other two samples ($O_{5.43}$ and $O_{5.48}$) we have also observed similar features corresponding to their FM $T_C$ (not shown here), possibly representing transitions arising from different clusters or domains co-existing in the sample. These features suggest the presence of inhomogeneties or electronic phase separation in $Eu_{0.9}Ca_{0.1}BaCo_2O_{5.5+\delta}$ similar to that in $LaBaCo_2O_6$ and other rare-earth cobaltites [27]. The ferromagnetic interactions are comparable to those responsible for the cluster glass behavior reported in some of the La$_{1-x}$Ca$_x$CoO3 cobaltites [27].

The temperature dependent resistivity of Eu$_{0.9}$Ca$_{0.1}$BaCo$_2$O$_{5.5+\delta}$ is shown in Fig. 6. Eu$_{0.9}$Ca$_{0.1}$BaCo$_2$O$_{5..48}$ ($\delta$ = -0.02) shows a clear metal-insulator transition (T$_{IM}$) around 350 K with an increase of the resistivity by about one order of magnitude. The $\delta$ = -0.28 and –0.07 compositions show semiconducting behaviour below the T$_{IM}$ and the low temperature resistivity is very high (Fig. 6). On the other hand, for $\delta$ = -0.02, 0.10 and 0.13 samples a large plateau in resistivity curves is observed below T$_{IM}$ to around 100 K, followed by a little increase on further lowering the temperature (Fig. 6). It is worthy to mention that in the calcium doped sample with $\delta$ = -0.02 the resistivity is about four orders less than that of the undoped EuBaCo$_2$O$_{5.5}$ phase at lowest temperature [1,7]. This result suggests that calcium doped samples with moderate oxygen content are much more conducting than the undoped sample. It is also clear that there are qualitative changes in the slope of these curves with the variation of oxygen stoichiometry in the ordered cobaltites.

To understand the nature of conduction mechanism for different phases of the compound we have carried out Seebeck coefficient, S(T), measurements below the I-M transitions. The thermopower measurement provides valuable information about the magnetic and electronic nature of charge carriers (hole/electron) which are absent in the magnetic and electrical resistivity studies. Moreover, S(T) data are less affected by the particle grains or grain boundaries, which often complicates the resistivity behaviour for polycrystalline samples. Fig. 7 shows the temperature variation S(T) for three different phases. It is observed that the S(T) value is positive for all the samples below the I-M transitions. With decreasing temperature the S(T) value increases and reaches a maximum (broad peak) value. The peak value for the samples $Eu_{0.9}Ca_{0.1}BaCo_2O_{5.22}$, $Eu_{0.9}Ca_{0.1}BaCo_2O_{5.48}$ and $Eu_{0.9}Ca_{0.1}BaCo_2O_{5.63}$ are 162, 175 and 115 K respectively. At lower temperature the S(T) value decreases continuously as expected for oxygen deficient 112-ordered cobaltites [5,13,22]. Since all the samples are poor conductors at low temperature, hence the data became unreliable at low temperature for few samples. It is important to notice that the change in slopes for $Eu_{0.9}Ca_{0.1}BaCo_2O_{5.48}$ and $Eu_{0.9}Ca_{0.1}BaCo_2O_{5.63}$ are higher than for the $Eu_{0.9}Ca_{0.1}BaCo_2O_{5.22}$ phase. Also the maximum value of thermopower for $Eu_{0.9}Ca_{0.1}BaCo_2O_{5.22}$ is the lowest, although resistivity is higher, among all three samples. At room temperature the highest thermopower value, 55 µV/K is obtained for $Eu_{0.9}Ca_{0.1}BaCo_2O_{5.48}$ and other two samples exhibit almost similar values around 8 µV/K.

**Discussion**

These results show that the variation of the oxygen content in this type of Ca-doped sample significantly affects the physical properties of 112-ordered cobaltites. First, we will discuss the magnetic behaviour above $T_C$, which covers the $T_{IM}$ region and consequently the controversial spin state transition of cobalt ions across $T_{IM}$. We have plotted $\chi^{-1}$ versus temperature for all the compositions as shown in Fig. 8 and data are fitted with Curie-Weiss law above and below the $T_{IM}$ temperature to obtain effective moment ($\mu_{eff}$). The antiferromagnetic interaction above $T_{IM}$ changes to a ferromagnetic one below $T_{IM}$ for $\delta > -0.02$. The variation of $\mu_{eff}$ per formula unit with oxygen content is shown in Fig. 9. The $\mu_{eff}$ values above $T_{IM}$ vary from ~3 to 4.25 $\mu_B$/Co in good agreement with the reported values [3]. However a large decrease in $\mu_{eff}$ (2 to 4 times) is observed below $T_{IM}$ (Fig. 9), which cannot be accounted for by considering the HS-LS transition scenario [6]. Thus, it is most likely that the metal-insulator transition is driven by the structural transition and does not involve any spin state transition, in agreement with the experimental and theoretical studies performed recently [28-32]. Such a structural transition leads to a change in the mode of coupling of the spins which is reflected by the crossover of an antiferromagnetic to ferromagnetic type interaction across the $T_{IM}$ for $\delta >$ -0.02, within the paramagnetic state. A rather similar mechanism was described for half doped manganites [33].

We will now discuss the effect of the oxygen content on this ferromagnetic phase. We have shown earlier that in the doped system, divalent calcium may locally change the crystal field and favour a partial and local disproportionation of trivalent cobalt $2Co^{3+}$ into $Co^{4+}$ and $Co^{2+}$ [24-26]. The $Co^{4+}$ ions contribute to $Co^{3+}$—O—$Co^{4+}$, ferromagnetic superexchange interactions, whereas HS $Co^{3+}$ and HS $Co^{2+}$ will form $Co^{2+}$—O—$Co^{3+}$ antiferromagnetic interactions [34]. This $Co^{3+}$—O—$Co^{4+}$ ferromagnetic interaction dominates over the competing antiferromagnetic interaction. It was also observed, that $T_C$ increases-with the calcium content though the average cobalt valence remains close to + 3. This was explained by the fact that the disproportioniation of cobalt around each Ca center creates a FM domain and that, consequently the ferromagnetism expands with the calcium content [24,25]. In the present case of $Eu_{0.9}Ca_{0.1}BaCo_2O_{5.5+\delta}$, calcium induces the presence of $Co^{4+}$ for $\delta$ close to zero, according to the dispropotioniation equation $2Co^{3+}$ $\leftrightarrows Co^{4+} + Co^{2+}$ (1). This is what is observed for $\delta$=-0.02. However as $\delta$ decreases, the $Co^{2+}$ content increases and displaces the equilibrium (1) towards the left (formation of $Co^{3+}$), decreasing then the $Co^{4+}$ content. Consequently, the $Co^{4+}$-O-$Co^{3+}$ FM interactions decrease as δ decreases from -0.02 to -0.28. Indeed, at 10 K the oxides with δ=-0.07 and δ=-0.28 are AFM, suggesting that the $Co^{4+}$ content may be very weak or negligible. Now, the equilibrium (1) is most probably sensitive to temperature, especially for δ values closer to zero, being displaced to the right, i.e favouring the disproportionation as temperature increases. This would explain the appearance of FM at 100 K for δ=-0.07, in contrast to AFM at the same temperature for δ=-0.28. Naturally, for δ>-0.02 an excess of $Co^{4+}$ is introduced, and there remains always enough $Co^{3+}$, to get $Co^{3+}$-O-$Co^{4+}$ interactions, leading to ferromagnetism.

In the present investigation we have two sources of $Co^{4+}$ ions: higher δ and disproportionation product in presence of divalent calcium. Thus, we would expect the higher $T_C$ in the samples with larger δ. But we have obtained the reverse one (inset of Fig. 3(b)). In the undoped sample $T_C$ decreases with the increase in $Co^{4+}$ content [21]. This result indicates that though the calcium doping destabilizes the low temperature antiferromagnetic state and expands the FM region, the generic magnetic feature of undoped $LnBaCo_2O_{5.5}$ is retained. Thus calcium substitution eventually leads to magnetic inhomogeneity or phase separation where ferromagnetic phases are embedded inside the antiferromagnetic matrix. It is expected that $Co^{4+}$ will sit preferentially in the octahedra with an IS state (S = 3/2), whereas $Co^{2+}$ will move to pyramids keeping a HS state (S = 3/2) leading to the gradual decrease in the effective moments with the increase in $Co^{4+}$ content (Fig. 9). Due to coexistence of FM and AFM phases at low temperature for some of the samples we have noticed unsaturated behavior of the M-H curve even at higher fields, which is a characteristic feature of glassy-ferromagnetic system. To confirm our assumption we have studied frequency dependent magnetic measurements and the obtained data for all samples are akin to glassy-ferromagnetic materials [27].

The change in resistivity with calcium doping can be understood on the basis of cobalt disproportionation. In the undoped compound with $\delta = 0$ the conventional semiconducting transport occurs, due to the mobility of the thermally excited hole and electron corresponding to the $Co^{4+}$ and $Co^{2+}$ species, deriving from ground state $Co^{3+}$ ions [6]. Calcium doping causes the formation of $Co^{4+}$ and $Co^{2+}$ which can easily transfer the corresponding hole and electron to the neighbouring $Co^{3+}$ ion, leading to the decrease in resistivity [24,25]. The doped carriers are strongly localized in the lower $\delta$ side ($\delta = 0.28$ and -0.07) and thus, the resistivity is high, whereas for higher $\delta$ values ($\delta \geq -0.02$) the carriers become slightly itinerant. The nature of the carrier is found to be p-type from the thermopower measurement for the whole range of $\delta$.

The obtained S(T) data signify that all three samples exhibit a semiconducting type behaviour of thermopower similar to 112-ordered cobaltites [5,13,22]. Moreover, the S(T) behaviour corroborates the trend of resistivity behaviour up to a certain temperature range, i.e. unusually below a certain temperature it decreases with lowering the temperature. This type of S(T) behaviour is unexpected for semiconducting thermoelectric materials due to trapping or localization of charge carriers at low temperature and the value should increase with decreasing temperature. It is also noted that for all three samples the S(T) value is positive throughout the measured temperature range, which implies p-type polaronic conductivity or hole like carriers in the samples. The unusual S(T) behaviour at low temperature has recently been explained by electron magnon scattering mechanism for 112-ordered cobaltite [22]. Our present data are similar to the reported cobaltite [13, 22], hence we expect that the mechanism will also be valid for the present systems.

**Conclusion**

The origin of ferromagnetic state in the calcium doped $Eu_{1-x}Ca_xBaCo_2O_{5.5+\delta}$ is the disproportionation. However, the generic feature of the magnetism of $EuBaCo_2O_{5.5}$ is retained even in the calcium doped phase as suggested by the investigation of oxygen disparity. We propose a phase separation scenario involving the formation of ferromagnetic domains embedded within the antiferromagnetic matrix, which is reflected by the various magnetic measurements. The transport measurements reveal the p-type nature of the carrier in all the samples. Finally the $T_{IM}$ is driven by the structural transition and is not related to any spin state transition.

**Figure Captions**

**Fig. 1** XRD patterns of $Eu_{0.9}Ca_{0.1}BaCo_2O_{5.50+\delta}$.

**Fig. 2** Variation of cell parameters as a function of $\delta$. Inset shows the change in cell volume with $\delta$.

**Fig. 3** M(T) curves of $Eu_{0.9}Ca_{0.1}BaCo_2O_{5.50+\delta}$ registered at an applied field of 100 Oe. The inset in (b) shows the $T_C$ variation with $\delta$.

**Fig. 4** M(H) curves of $Eu_{0.9}Ca_{0.1}BaCo_2O_{5.50+\delta}$ registered at (a) 10 K and (b) 100 K.

**Fig. 5** Frequency dependent ac-susceptibility plots as a function of temperature for $Eu_{0.9}Ca_{0.1}BaCo_2O_{5.63}$.

**Fig. 6** Temperature dependent resistivity of $Eu_{0.9}Ca_{0.1}BaCo_2O_{5.50+\delta}$. The filled up triangle (▲) data for $\delta = 0.02$, taken from Ref. 21, is shown for comparison.

**Fig. 7** The temperature dependent thermopower for $Eu_{0.9}Ca_{0.1}BaCo_2O_{5.50+\delta}$.

**Fig. 8** Inverse susceptibility, $\chi^{-1}$ plots as a function of temperature for $Eu_{0.9}Ca_{0.1}BaCo_2O_{5.50+\delta}$.

**Fig. 9** The change in effective magnetic moments both above and below $T_{IM}$ per formula unit of $Eu_{0.9}Ca_{0.1}BaCo_2O_{5.50+\delta}$ as a function of oxygen content.

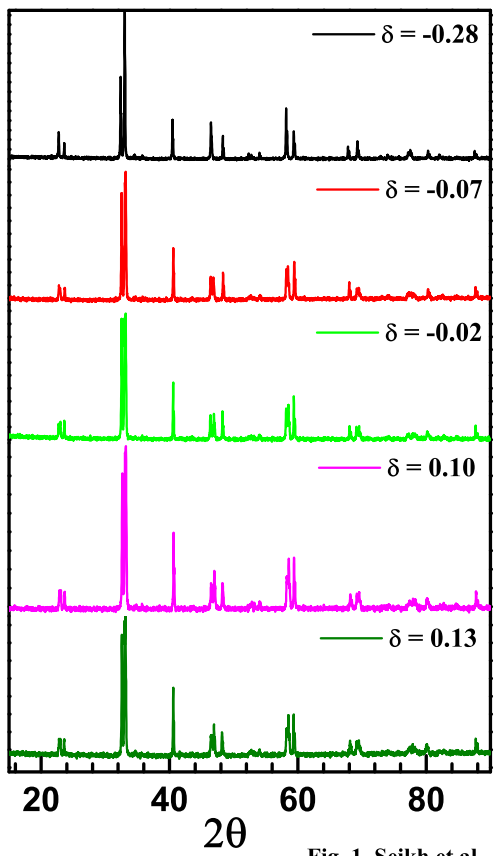

Fig. 1, Seikh et al.

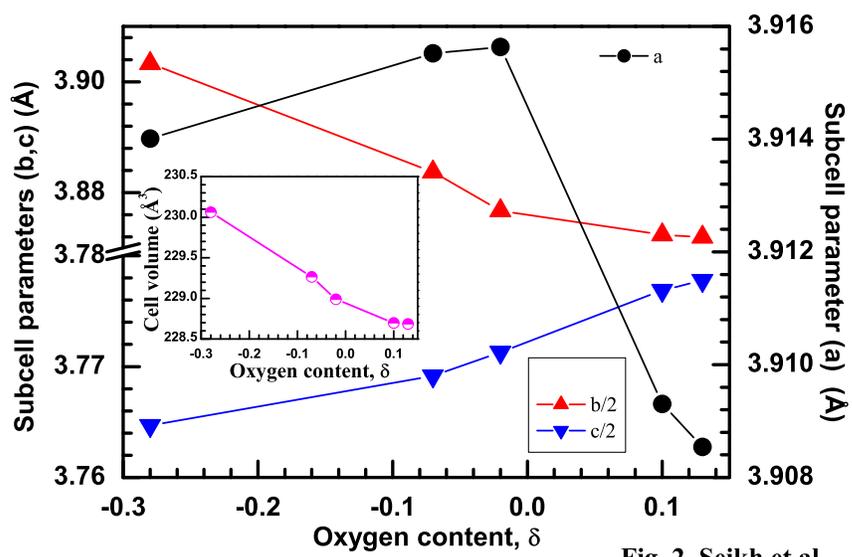

Fig. 2, Seikh et al.

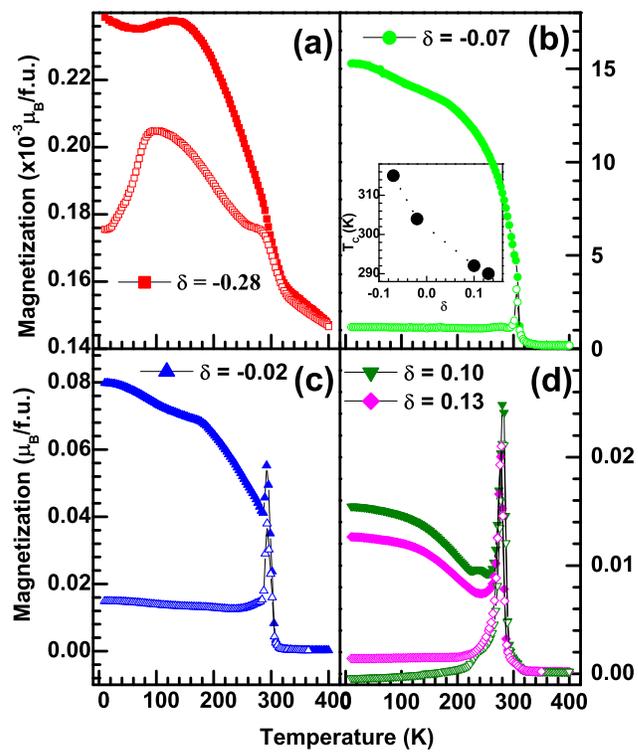

Fig. 3, Seikh et al.

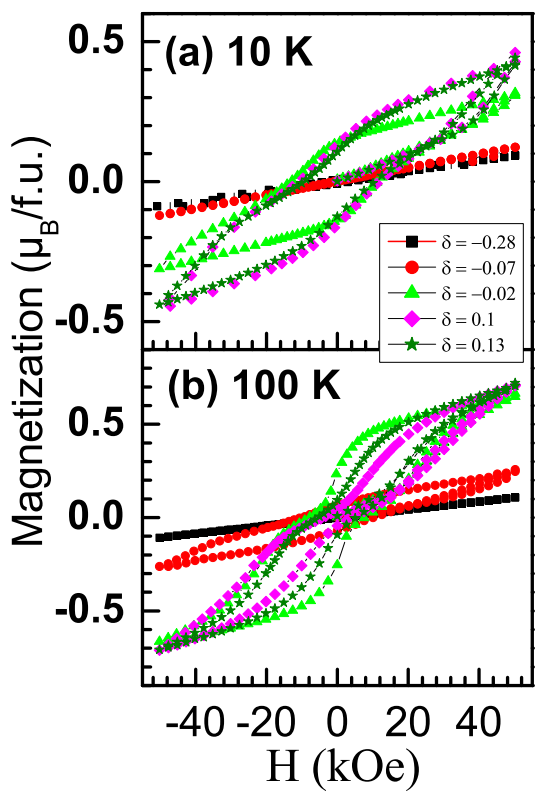

Fig. 4, Seikh et al.

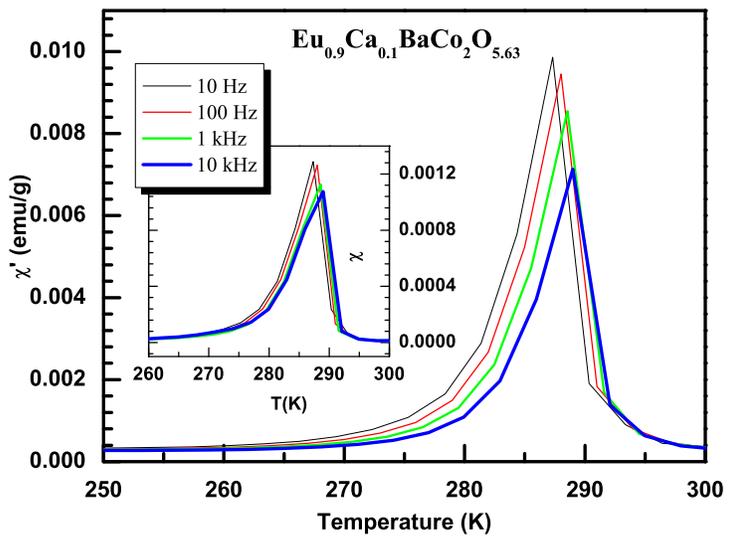

Fig. 5, Seikh et el.

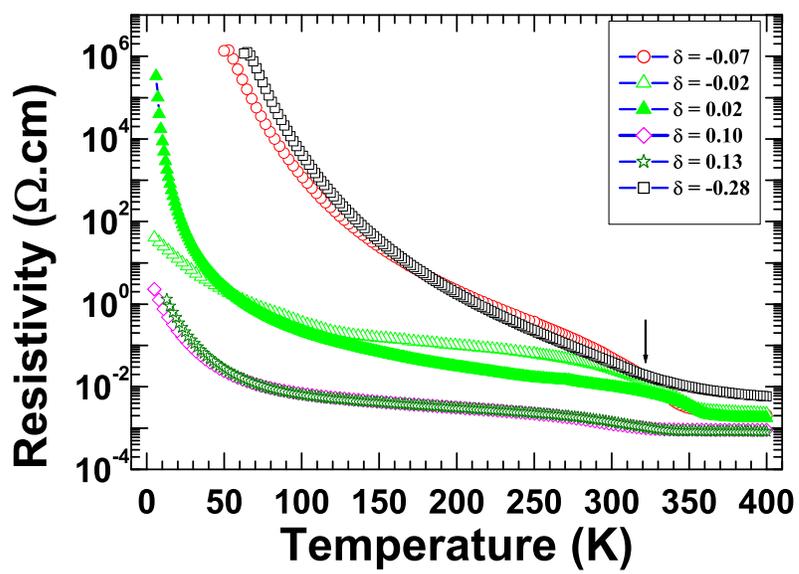

Fig. 6, Seikh et al.

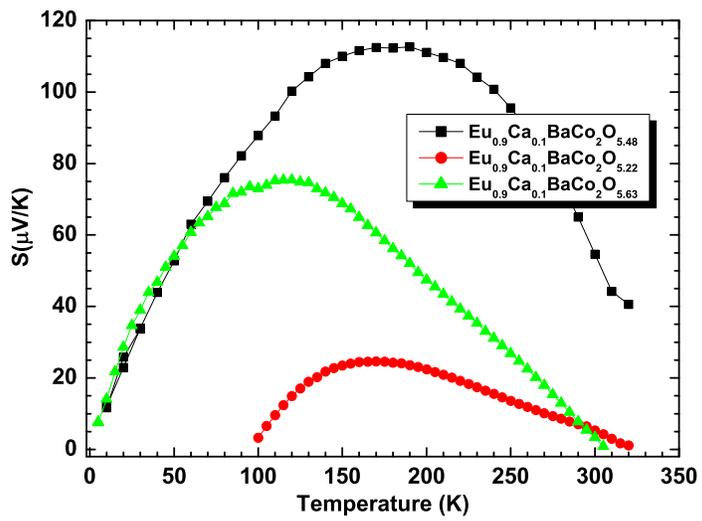

Fig. 7, Seikh et al.

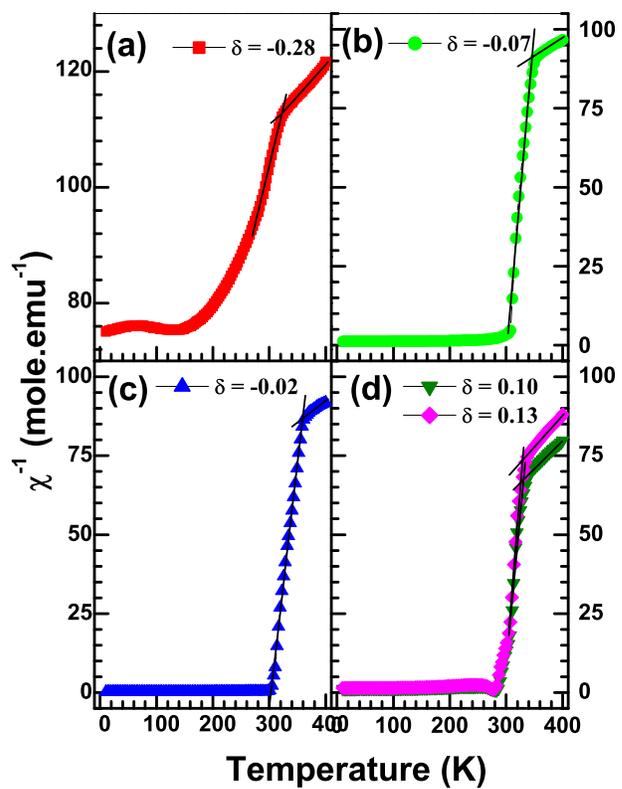

Fig. 8, Seikh et al.

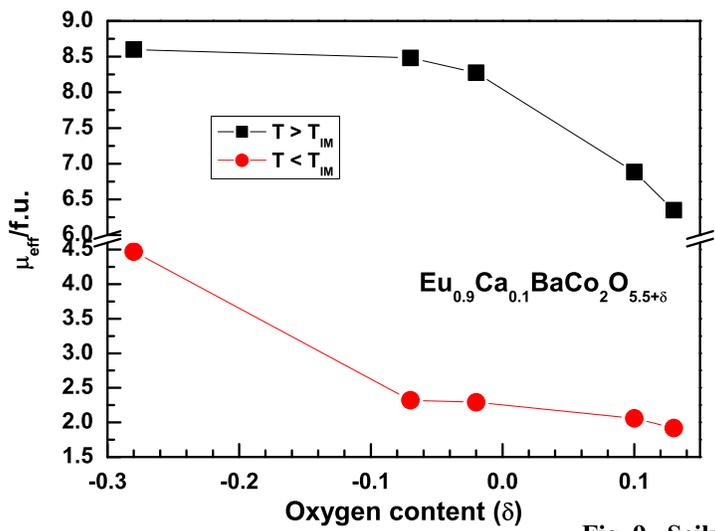

Fig. 9, Seikh et al.